\documentclass[12pt,preprint,usenatbib]{emulateapj}

\newcommand{\change}[1]{{#1}}

\newcommand{\teff}{$T_{\mathrm{eff}}$}
\newcommand{\muhz}{$\mu$Hz}
\newcommand{\numax}{$\nu_{\mathrm{max}}$}
\newcommand{\dnu}{$\Delta\nu$}

\shorttitle{Oscillations in K giants with the HST}
\shortauthors{Stello \& Gilliland}

\begin{document}


%\title{Detection of stellar oscillations in metal-poor globular cluster
%  with the HST}  
%\title{Detection of stellar oscillations in K giants in metal-poor
%  cluster with the HST}  
%\title{Solar-like oscillations in metal-poor K giants with the HST}  
%\title{Solar-like oscillations in metal-poor cluster giants with the HST}  
\title{Solar-like oscillations in a metal-poor globular cluster with the HST\altaffilmark{*}}  
%\title{Red giant oscillations from HST time series on NGC 6397}  
%\title{Stellar oscillations in metal-poor globular cluster
%  with the HST}  
%\title{HST detects solar-like oscillations in metal-poor globular
%        cluster NGC6397}
%\title{A search for solar-like oscillations in the metal-poor globular
%        cluster NGC6397 with HST}
%\title{HST detects oscillations in K giant population of the metal-poor globular
%        cluster NGC6397}


\author{Dennis Stello\altaffilmark{1}, 
  Ronald L. Gilliland\altaffilmark{2}} 

\altaffiltext{1}{Sydney Institute for Astronomy (SIfA), School of Physics,
  University of Sydney, NSW 2006, Australia; stello@physics.usyd.edu.au.} 
\altaffiltext{2}{Space Telescope Science Institute, 3700 San Martin Drive,
  Baltimore, Maryland 21218, USA; gillil@stsci.edu.}

\clearpage

\altaffiltext{*}{Based on observations with the NASA/ESA Hubble Space
  Telescope, obtained at
the Space Telescope Science Institute, which is operated by the Association
of Universities for Research in Astronomy, Inc., under NASA contract NAS
5-26555.}

\begin{abstract}
We present analyses of variability in the red giant stars in the
metal-poor globular cluster NGC6397, based on data
obtained with the {\em Hubble Space Telescope}.
%OR We present the detection of oscillations from a population og K giants in the
%metal-poor globular cluster NGC6397, based on data 
%obtained with the Hubble Space Telescope over a 23-day period.
We use an non-standard data reduction approach to turn a 23-day
observing run originally aimed at imaging the white dwarf population, into
time-series photometry of the cluster's highly saturated red giant stars. 
With this technique we obtain noise levels in the final power spectra
down to 50 parts per million, which allows us to
search for low amplitude solar-like oscillations.
We compare the observed excess power seen in the power spectra with estimates
of the typical frequency range, frequency spacing and amplitude from
scaling the solar oscillations. We see evidence that the detected
variability is consistent with solar-like oscillations in at least one
and perhaps up to four stars. 
With metallicities two orders of magnitude
lower than of the Sun, these stars present so far 
the best evidence of solar-like oscillations in such a low metallicity environment.

\end{abstract}


\clearpage

\section{Introduction} 

Theoretical models predict that cool stars with convective envelopes
exhibit oscillatory motion due to excitation of a large 
number of eigen modes by the turbulent gas motion near the surface. 
The resulting p-mode oscillations are standing sound waves with the
restoring force being pressure, hence their name, which therefore depend
on the internal structure of the star. For historical reasons these
oscillations are also known as solar-like oscillations. The technique of
analyzing these oscillations to infer and constrain
the stellar properties, called Helioseismology when applied to the Sun, has 
%been developed OR
shown some remarkable results 
over the last few decades, and is starting to 
%take off OR
show promising results 
when applied to other stars -- known as asteroseismology,
e.g. see \citet{BeddingKjeldsen07} and references therein.

%maybe rewrite next two lines depending on what we need to say re out
%analysis in the paper
A typical characteristic of p-mode oscillations is an almost equidistant
frequency spacing between modes of successive radial order with a broad,
smoothly varying amplitude profile as a function of frequency. 
Amplitudes are only a few tens of centimeters per second in velocity and
a few parts per million (ppm) in intensity in stars similar to the Sun,
while the more vigorous convection in red giant stars
generate amplitudes much larger, reaching up to several meters per second
or the order of 1000 ppm ($\sim$1mmag) in intensity. 
%remove next sentence if make intro more general solar-like
The higher signal-to-noise and
prospects of obtaining additional measures to constrain the stellar
parameters at this interesting evolutionary state makes red giants obvious
targets for asteroseismic
study.

%This par. sets the scene for our use of the scaling relations for \numax
%and \dnu, and the test of the two amplitude scaling relations
A few simple relations exist to predict the amplitude, frequency, and
frequency spacing of p-mode oscillations. 
There is quite strong observational evidence that the typical frequency
range of the oscillations scales with the acoustic cut-off
frequency \citep{KjeldsenBedding04,Stello08}, as suggested by \citet{Brown91}. There
is also good agreement between the observed 
frequency spacings and those predicted by scaling the square root of
the stellar density \citet{KjeldsenBedding95,Kjeldsen08a}. 
However, there is less of an agreement when it comes to the amplitude of
the modes, which might be related to the often unknown mode
lifetime. \citet{KjeldsenBedding95} suggested that the amplitude per mode 
would be proportional to the luminosity-to-mass ratio in velocity and 
$LM^{-1}$\teff$^{-2}$ in intensity. This was based on measured
amplitudes from various types of oscillating stars.
More recent theoretical studies by \citet{Samadi05} indicate that
$(L/M)^{0.7}$\teff$^{-2}$ in intensity might be a better scaling factor.

%This paragraph maybe move up
{\change Oscillations have now been clearly detected in a few bright red giant field
stars, both in velocity \citep{Frandsen02,Ridder06,HatzesZechmeister07} and
photometry \citep{Barban07}.} 
While earlier datasets were not adequate to clearly establish the p-mode
nature of the detected variability, the more recent measurements leave
little or no doubt as the reality of p-mode oscillations in red giants.
However, there is still no clear consensus on the mode lifetimes, and it is
still not well known to what extent non-radial modes are present. 
{\change For the
star $\xi\,Hya$ \citet{Stello06} found a mode lifetime of a few days
assuming only radial modes, which was significantly shorter than 
the theoretical value of about 17 days derived by
\citet{HoudekGough02}. Similarly for $\epsilon\,Oph$, \citet{Barban07}
also found a short mode lifetime assuming only radial modes, while
\citet{Kallinger08} used the same photometry to conclude that this star
showed both radial and non-radial pulsations with mode lifetimes of roughly
10-20 days.  }

After indications of p-mode oscillations were obtained in a large sample of
red giant stars from a 38-h run on the globular cluster 47 Tuc
([Fe/H]$\simeq-0.7$) with the {\em Hubble Space Telescope} (HST)
\citep{EdmondsGilliland96}, large efforts have been made to  
detect p-mode oscillations in stellar clusters based on extended
ground-based photometry. The open cluster M67 
([Fe/H]$\simeq0.0$) has so far only provided marginal detections in a few
red giants \citep{Gilliland93,Stello07}. The results by \citet{Stello07}
showed slightly better agreement with the $LM^{-1}$\teff$^{-2}$ scaling
than the $(L/M)^{0.7}$\teff$^{-2}$ scaling. However, no strong conclusions
could be drawn because the extent to which non-radial modes were present
could not be established.
Aimed at the red giant population of the metal-poor globular cluster M4
([Fe/H]$\simeq-1.2$) \citet{Frandsen07} were only able to establish upper
limits on the amplitudes, which were not in agreement with the
$LM^{-1}$\teff$^{-2}$ scaling, but still consistent with the more modest
estimates from the $(L/M)^{0.7}$\teff$^{-2}$ scaling. 
\citet{Frandsen07} suggested the lower amplitudes in M4 relative to M67
could be due to its lower metallicity.
This is supported by the velocity measurements of the subgiant field star
$\nu\,$Ind by \citet{Bedding06}, which is the most metal-poor star
([Fe/H]$\simeq-1.4$) for which p-mode oscillations have been 
detected so far.

In this paper we aim to detect p-mode oscillations from a group of red
giants in the metal-poor globular cluster NGC6397 ([Fe/H]$\simeq-2.0$;
\citet{Richer08}) from 23 days of photometry from the HST, and use those
results to test the scaling relations for the mode amplitudes.
First, we give a detailed description of the non-standard data reduction
that was required for this investigation in \S~\ref{observations}. Then, in  
\S~\ref{parameters} we provide the derivation of stellar parameters and
estimated asteroseismic characteristics, which leads to the time-series
analysis in \S~\ref{analysis}, and estimates of the mode amplitudes in
\S~\ref{amplitudes}. %followed by...
In \S~\ref{autocorr} we search for a regular pattern in the power spectrum of the
times series, and discuss our results in \S~\ref{discussion}. 
Finally, we give the conclusions in \S~\ref{conclusions}.

%Over the last decade or two there have been several attempts to detect p-mode
%oscillations in red giants. The most revisited case is Arcturus
%\citep{Smith87,Cochran88,Belmonte90,HatzesCochran94,Retter03} where there is
%general agreement between the detected periodicities of 2--3 days with the
%expected signal from p-mode oscillations 
%However, it was not untill the the refinements of the radial velocity
%techniques developed to persue extra solar planet detection that we saw the
%first clear detections 
%Only in a few cases a firm detection has been established.
%Driven by this (the prospect for seismo on RGs)

%\citep{Frandsen02} succeeded in obtaning a firm detection of
%p-mode oscillations in the G8 giant $\xi\,$Hya through velocity
%measurements. Subsequenctly, \citet{Barban04} and citet{Ridder06} reported
%detection of oscillations in two other red giants, $\eta\,$Ser and
%$\epsilon\,$Oph -- also in velocity. Recently, \citet{Barban07} showed the
%first clear photometric detection p-modes in a red giant.
%
%Intro all the M67 campaigns, and the M4 campaign.
%Intro RG papers (see my previous papers for refs plus the new stuff from
%Corot if public when we submit).

\section{Observations of NGC6397}\label{observations}
\begin{figure}
%\epsscale{1.0}
%\plotone{f1.ps}
\includegraphics{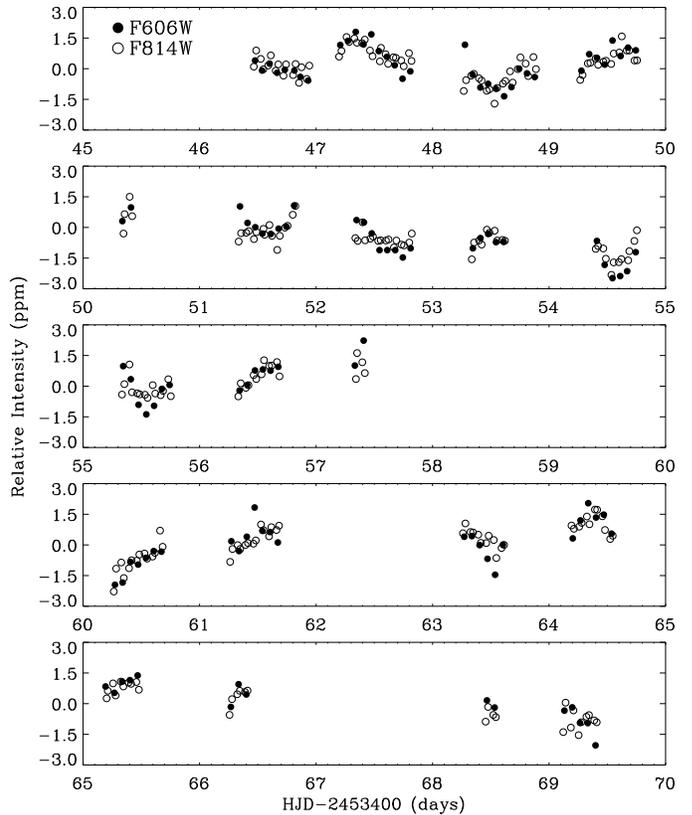}
\caption{Time series of the brightest cluster member (No 1) on the WFC1
  chip.  During this 23 day period 37\% of all HST orbits had data for 
  this project, and within orbits 38\% duty cycle was acheived.
\label{f1}} 
\end{figure}

%The field observed was located at 5 arcmin from the cluster center.
The data on the red giants presented in this paper were serendipitously
obtained as part of GO-10424 \citep{Richer08}, a program executed
over 2005 March 17 -- April 9 during 126 orbits with the HST's ACS/WFC
(Cycle 13). While the original program   
aimed to probe the bottom of the white dwarf cooling sequence of NGC 6397 
through deep imaging of a single field in this nearby globular
cluster \citep{Hansen07}, we use the 363 exposures %--
%one third through the F606W filter and two thirds through the F814W filter --
as one 23-day times series to allow detection of stellar
variability.
%Typically two exposures through the F814W filter and one through the F606W
%filter were obtained during each 96-minute orbit. 
During each 96-minute orbit, typically one exposure through the F606W
filter was obtained, which was 
%bracketed by two exposures through the F814W 
%filter separated by roughly 16 minutes. 
bracketed by two exposures through the F814W filter separated by roughly 16
minutes. Figure~\ref{f1} shows the time series of one of the stars.
The exposure times
were on average 713 seconds for F606W and 742 seconds for F814W.
However, the data acquisition was not optimized for our purpose,
the red giants were saturated, and images were dithered over several
pixels %($\pm10$ pixels in $x$, and $\pm3$ pixels in $y$) %in order to best
%average out flat-field irregularities, and pixel defects for combined
%imaging 
-- both effects adding noise to the final time series. 

%To correct for the dithering we estimated frame-to-frame offsets by fitting  
%PSFs to 42 non-saturated stars in each image, and solved for offsets,
%rotation, and plate scale changes for each frame relative to an overall
%mean. 
Due to saturation we had to abandon PSF fitting for the target stars, and
instead use aperture photometry.  
The goal of the following step was to find an aperture mask, as small as
possible, for each star that would always encompass all of the pixels that were
ever bled into for the over-saturated stars. 
We first formed a stack of the longest exposures after shifting them to a
common $x,y$ position. 
The frame-to-frame offsets were found by fitting PSFs to 42 non-saturated
stars in each image, and solving for offsets, rotation, and plate scale
changes relative to an overall mean.
Then, for each pixel in the stack we looked at the
distribution of counts, and assigned the pixel to be part of the aperture if
there was more than one value in the distribution near the known
saturation threshold of the chip. In case there was only one value near
saturation it would more likely be due to a cosmic ray event, and was ignored to
avoid spuriously large apertures. This was done separately for the two
filters, and was based on 84 exposures of 769 seconds in F606W
and 93 exposures of 804 seconds in F814W, the longest exposure time used for either filter.
We were now able to extract simple aperture photometry for the 510
brightest stars on chip WFC1 that did not bleed off the edges. This
included 255 stars that were slightly saturated (bleeding never exceeded a
radius of ~3 pixels), and 255 stars with significant saturation. 
For the brightest star the aperture contained
slightly more than 10,000 pixels and was nearly 1700 pixels
tall along the bleeding direction.
Due to small number statistics only one of the two chips -- WFC1 -- contained stars suitable for
this investigation which is aimed at red giants with the 
F814W magnitude $\lesssim$ 13.
		    %detection and not bleeding off the ccd edge

To reduce the spurious noise introduced by the dithering we decorrelated
the time series with a large number 
of external parameters.  For previous experiments with HST data where the
dithering had been within $\pm0.5$ pixels, the primary decorrelation
vectors were simply the $x,y$ offsets frame-to-frame \citep{Gilliland08}.
For these observations the dithering placed equal
amounts of data in 10 widely separated pixels, with data
points within each position uniformly distributed in time.
In examining the raw time series it was obvious that 
unique offsets in the photometry often existed for one (or more)
of these offset positions.  
We remove those offsets by decorrelating with nine independent vectors
while adjusting to a common zero point corresponding to the 10th dither
position.
For most time series linear correlation coefficients remained
small for most terms.  But, sometimes when photometric offsets unique to 
a dither position existed, the correlation could be larger than 0.9. 

Within each dither position there were additional sub-pixel offsets which
varied up to $\pm0.25$ pixels. These tended to be along an $x,y$
diagonal. Hence we formed 10 additional vectors -- one for each dither
position -- to remove photometry variations that correlated with the
sub-pixel offsets. Again, it was usually the case that linear correlations
between these sparsely populated vectors and the photometry were small. 
%, but sometimes they could be large if a star fell near a bad pixel and
%moved with respect to it.

We included a total of four more decorrelation vectors: (1) a median over
the 510 input stars to enforce an ensemble mean;
(2) the local sky value frame-to-frame to correct any systematic errors in
the initial sky subtraction; (3) the frame-to-frame exposure time to
correct non-linearities; (4) and a goodness of fit parameter from the
earlier step of fitting PSFs to deal with any focus changes.
Again, most correlations remained small, but occasionally a star would
see dramatic improvement from one of these terms.
The decorrelations were performed iteratively with allowance
for dropping points for which deviations were larger than $4\sigma$.
Over the 242 frames in F814W and 121 frames
in F606W an average of 4.4 points and 1.4 points were 
removed, respectively.

To test that the large number of decorrelation parameters (24 in total) did
not wreak havoc with the data, we performed the following experiment.
Starting with the original non-decorrelated time
series for the 510 stars we injected sinusoidal signals 
with amplitudes of 0.001 and frequencies that uniformly spanned 1 to
100\muhz, after which we decorrelated the resulting time
series. Separately, we
injected the same set of sinusoids in the original data after
decorrelation, and then compared signals between the two. 
This comparison showed that our decorrelation process made almost no
difference for the injected sinusoids at any frequency.  

\begin{figure}
%\epsscale{1.0}
\includegraphics{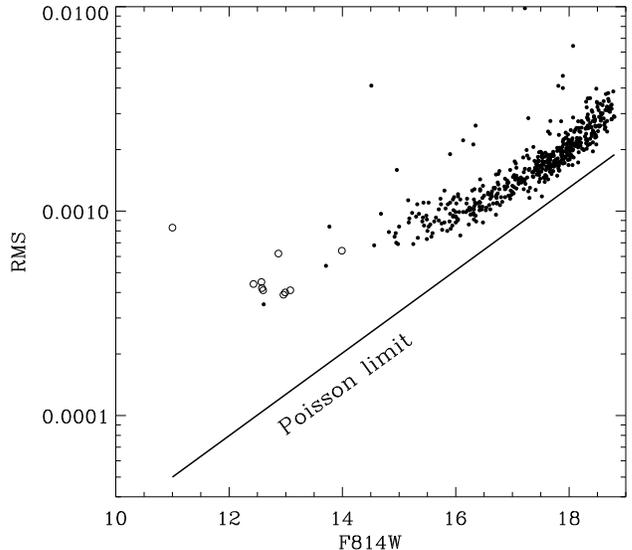}
\caption{Standard deviation of time series as a function of magnitude for
  all 510 stars examined. Open circles show the 10 brightest cluster
  members listed in Table~\ref{tab1}. All 510 stars are saturated, which
  explains their rms offset relative to the Poisson limit marked by the
  solid line. 
 \label{f1a}} 
\end{figure}

Overall, the final time series rms remains about 50\% above
the Poisson limit in the stars that are slightly saturated,
to a factor of two above for more strongly saturated stars, as illustrated
in Figure~\ref{f1a}.
The inability to eliminate most cosmic rays in these heavily
saturated stars can explain much of this excess residual.
Among the red giant stars we see additional excess rms, which is likely
intrinsic stellar variability.

\section{Stellar parameters}\label{parameters}
\begin{figure}
%\epsscale{1.0}
\includegraphics{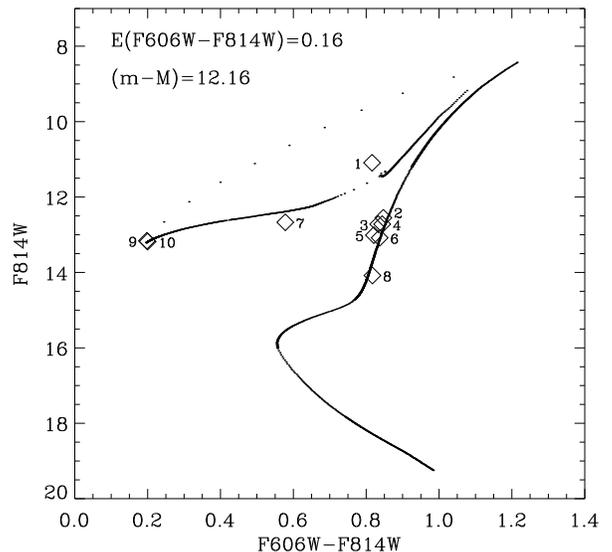}
\caption{Color-magnitude diagram of bright stars in NGC6397, and an
  overlaid isochrone from the BaSTI grid (see text). 
\label{f2}} 
\end{figure}

%Maybe intro sentence here
Within the 510 stars we used Table 1 of \citet{Richer08} to select all of
the proper motion cluster members brighter than 14.1 magnitude.
Adopting the absolute photometry from \citet{Richer08} we show in
Figure~\ref{f2} those bright cluster members and an appropriate 
isochrone from the BaSTI grid \citep{Pietrinferni04} for age = 13 Gyr,
$Y = 0.245$, and $Z=0.0001$ corresponding to [Fe/H]$=-2.3$. The model uses
scaled solar mixture, no overshooting, and assumes a mass loss parameter of
$\eta=0.4$.
To match the bright cluster stars with the isochrone %maybe remove this line
we adopted a reddening and distance modulus of $E(F606W-F814W)=0.18$ and
$(m-M)_\mathrm{F814W}=12.6$ respectively in agreement with previous values
in the literature (e.g. see \citet{Richer08} and references therein).
We obtained the stellar parameters $L/$L$_\odot$, \teff\, and $M/$M$_\odot$  by 
%(1)
matching each star to the isochrone. 
%(2)
%locating the point along the isochrone with the smallest deviation in
%brightness and color. 
We then estimated the frequency at maximum power, the characteristic
frequency spacing between modes of successive radial order, and the
mode amplitude using the scaling relations from
\citet{KjeldsenBedding95} and \citet{Samadi05}.
The results are listed in Table~\ref{tab1}. The first column is the star
number, second to fifth columns are $x$, $y$, F814W, and F606W-F814W from
Table 1 of \citet{Richer08}, columns 6--9 are luminosity,
temperature, radius and mass, and the last four columns are expected frequency of
maximum power, the characteristic frequency spacing, and
the amplitude based on two different scaling relations.
%And the last two columns are time series rms in the
%F814W and F606W cases separately.

\begin{table*}
%{\footnotesize
\begin{center}
\caption{Stellar parameters\label{tab1}}
%\begin{tabular}{r|r@{(}lrr|rr@{(}l|rr|rr}
\begin{tabular}{lllllllllllll}
\tableline\tableline
     &     &      &       &             &               & \teff &               &               & \numax\  & \dnu\    & $L/M$   & $(L/M)^{0.7}$\\
Star & $x$ & $y$  & F814W &F606W$-$F814W& $L/$L$_\odot$ & (K)   & $R/$R$_\odot$ & $M/$M$_\odot$ & (\muhz)  & (\muhz)  & (ppm)   & (ppm)      \\
 (1) & (2) & (3)  &  (4)  &   (5)       &    (6)        & (7)   &  (8)          &  (9)          & (10)     & (11)     & (12)    & (13)     \\
\tableline
1 & 1824.0&  922.0& 11.09 & 0.816       &  174          &  5100 &  16.9      &  0.72          &  8.31      &  1.65   &  983   &  190 \\
2 & 1881.0& 1181.0& 12.55 & 0.847       &  46.7         &  5192 &  8.46      &  0.79          &  35.9      &  4.86   &  234   &  68.7 \\
3 & 552.1 &  992.0& 12.72 & 0.833       &  39.8         &  5219 &  7.73      &  0.79          &  42.8      &  5.56   &  197   &  60.8 \\
4 & 2186.0& 1794.0& 12.72 & 0.844       &  39.8         &  5219 &  7.73      &  0.79          &  42.8      &  5.56   &  197   &  60.8 \\
5 &  923.2&  937.0& 13.01 & 0.821       &  30.8         &  5262 &  6.69      &  0.79          &  57.1      &  6.92   &  150   &  49.9 \\
6 & 2097.0& 1827.0& 13.08 & 0.837       &  28.8         &  5274 &  6.44      &  0.79          &  61.6      &  7.33   &  139   &  47.4 \\
7 & 2254.0&  372.0& 12.67 & 0.578       &  58.4         &  7108 &  5.05      &  0.72          &  79.1      &  10.1   &  170   &  45.5 \\
8 & 1520.0&  909.0& 14.08 & 0.817       &  11.7         &  5421 &  3.89      &  0.79          &  170       &  15.6   &  53.9  &  23.9 \\
9 & 1310.0& 1506.0& 13.16 & 0.198       &  48.8         &  8931 &  2.92      &  0.72          &  210       &  22.9   &  90.0  &  25.4 \\
10& 2049.0& 1549.0& 13.16 & 0.200       &  48.8         &  8931 &  2.92      &  0.72          &  210       &  22.9   &  90.0  &  25.4 \\
%     &       &             &           & Teff&           & Irms&  Vrms \\
%Star & F814W & F606W-F814W & L/L_\odot & (K) & R/R_\odot & Irms&  Vrms \\
%508& 11.09&   0.816&    145.0&4917.&16.66& .00083& .00107 \\
%355& 12.72&   0.833&     33.1&5219.& 7.06& .00044& .00062 \\
%510& 12.55&   0.847&     38.5&5192. &7.70& .00044& .00055 \\
%457& 12.72&   0.844&     33.1&5219.& 7.06& .00041& .00041 \\
%275& 12.67&   0.578&     34.6&5211.& 7.24& .00042& .00034 \\
%344& 13.01&   0.821&     25.4&5266.& 6.08& .00062& .00042 \\
%467& 13.08&   0.837&     23.9&5277.& 5.87& .00039& .00052 \\
%421& 13.16&   0.198&     22.2&5289. &5.63 &.00040& .00032 \\
%509& 13.18&   0.200&     21.8&5293.& 5.57& .00041& .00035 \\
%333& 14.08&   0.817&      9.6&5429.& 3.52& .00071& .00075 \\
%355    552.1  992.0 12.72   .833    .07  -.24
%344    923.2  937.0 13.01   .821    .00  -.35
%421   1310.0 1506.0 13.16   .198   -.27  1.27
%333   1520.0  909.0 14.08   .817    .14  -.07
%508   1824.0  922.0 11.09   .816    .51   .10
%510   1881.0 1181.0 12.55   .847    .01  -.17
%509   2049.0 1549.0 13.18   .200    .10  -.21
%467   2097.0 1827.0 13.08   .837    .07  -.44
%457   2186.0 1794.0 12.72   .844    .20  -.63
%275   2254.0  372.0 12.67   .578   1.17  -.18
\tableline
\end{tabular}
%\vspace{-1cm}
%% Any table notes must follow the \end{tabular} command.
\tablenotetext{}{(2)--(5) are from Table 1 in \citet{Richer08}}
\tablenotetext{}{If mass loss is ignored all the stars would have a mass of
  roughly 0.79 M$_\odot$.}
\end{center}
%}
\end{table*}

%With these parameters in hand I can predict what oscillations
%should be based on the results in my recent Gilliland 2008
%AJ, 136, 566 paper.  The predictions are:
%Star 508  Amp $\sim700$ ppm peak near 10\muhz.
%Stars 355, 510, 457  Amp $\sim100$ ppm with peak near 50\muhz.

\section{Time-series analysis}\label{analysis}
To enhance the signal-to-noise we sought to combine the data obtained through
the two filters.  
Fortunately, there are no phase changes in the solar oscillations observed
in different filters with bandpass differences of roughly 200 nm
\citep{Jimenez99}, and we therefore expect the same adiabatic behaviour for
high-order solar-like oscillations in other stars. Hence, we combine the
data by simply scaling the amplitude of one of the datasets to accommodate
the $1/\lambda_{\mathrm{filter}}$ dependence while adjusting
their weights to preserve the correct noise level. 
%We used uniform weights for data obtained in the same color band. 
%(see Gilliland2008 table 3 for empirical ratio between the two...it
%seems that the ratio depends on teff and Lum...but is it just a results of
%S/N being dependent on teff and Lum so that a high S/N (signal dominated)
%gives higher ratio while low S/N (noise dominated) gived ratio~1?) 

Out of the full set of 510 stars power spectra are generally
flat, with little evidence of excess low frequency noise
coming in as would always be the case with ground-based data.
\begin{figure}
%\epsscale{1.0}
\includegraphics{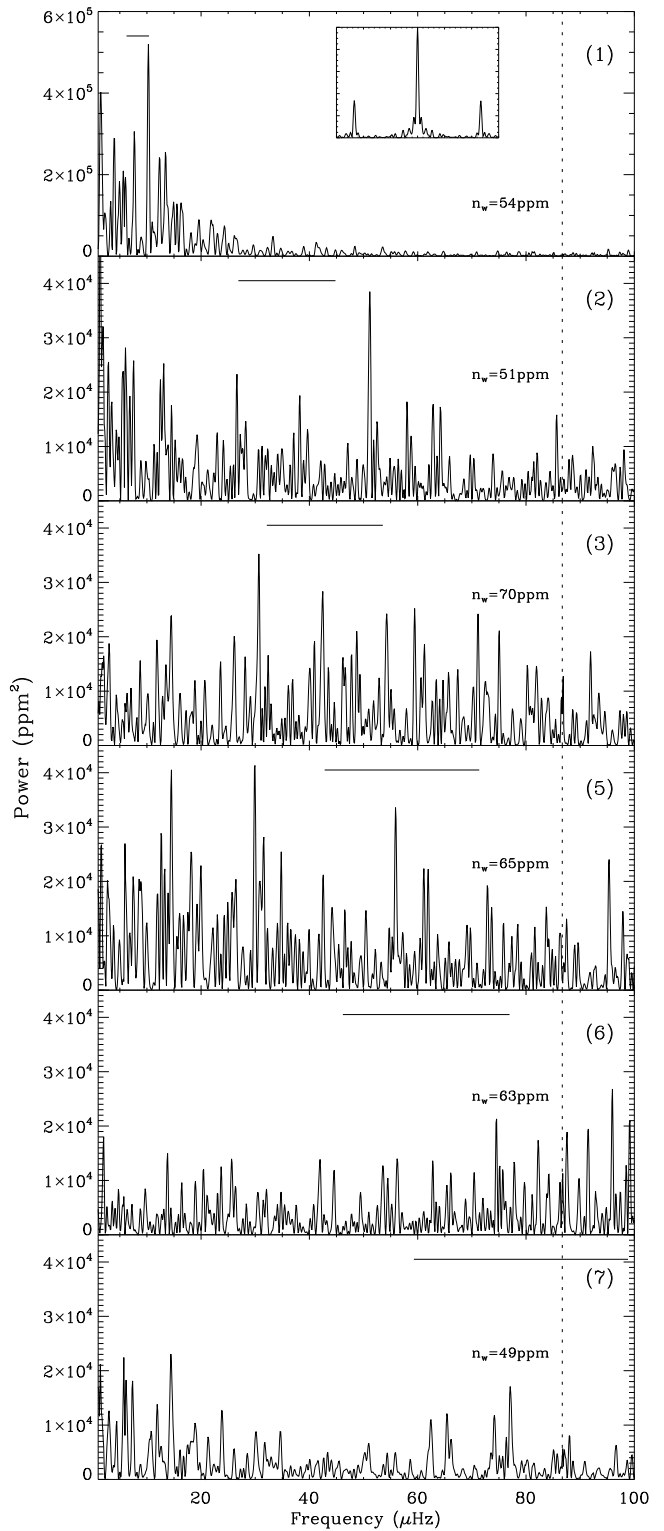}
\caption{Fourier spectra of red giant stars listed in Table~\ref{tab1}. The
  dotted lined indicates half the orbital frequency, and
  $n_{\mathrm{w}}$ is the mean level in the spectra in the range
  67--87\muhz\ converted to amplitude. The expected frequency range of
  the oscillations (\numax$\pm \frac{1}{2}$\numax) is marked by the
  horizontal bar. The inset in the top panel shows the 
  spectral window on the same frequency scale as the main panel, and is
  representative for all stars. Star numbers relating 
  to Table~\ref{tab1} are marked in each panel. 
\label{f3}} 
\end{figure}
Of the 10 red giants listed in Table~\ref{tab1} we show power spectra
(weighted Fourier transform) of the best candidates in Figure~\ref{f3}. These
were selected to have a white noise level below 100ppm (in amplitude) and with
expected oscillation frequencies below 100\muhz\ to avoid strong aliasing
from the orbital frequency of the spacecraft. 
%Stars No. 8, 9, and 10 all had expected amplitudes too low to be detected and
%with frequencies near the satellite orbital frequency, which would 
%hamper the analysis due to strong aliasing.
%REMOVE SMOOTHED SPEC FROM THIS PLOT
%with a smoothed version overlaid to enhance visability of envelopes of
%excess power. 
Due to daily gaps in the data the spectral window is similar to
ground-based single-site observations (see inset in top panel).

A quite strong excess of power at low frequencies is seen for star No. 1
consistent with the expected frequencies from solar-like oscillations
indicated with a horizontal line. No other 
star in our sample shows such a steep increase in power towards low
frequencies.
Note the much larger ordinate range in the top panel.
However, the high peaks in this frequency range are only just resolved and
do not support a detailed frequency analysis (see \S~\ref{autocorr}).

\begin{figure}
%\epsscale{1.0}
\includegraphics{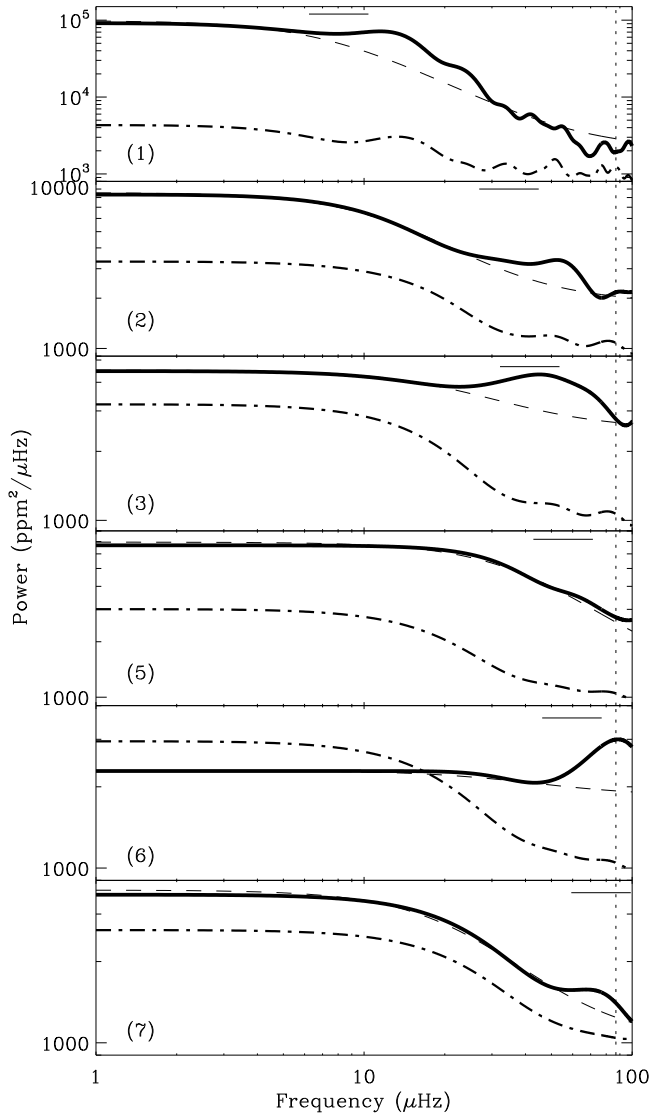}
\caption{Power density spectra of the stars shown in Figure.~\ref{f3}
  convolved with a Gaussian with width 4\dnu. With the same degree
  of smoothing we show for comparison the average smoothed spectrum of two
  stars where no detectable oscillations are expected (dash-dot). 
  The dashed line shows a representative one-component Harvey model
  including a white noise term as shown in equation~\ref{harvey}. 
\label{f4}} 
\end{figure}

The rest of the sample show at most a few intriguing peaks.
%One star hints the existence of excess power near 87\muhz. However,
%this is half the orbital frequency and we are therefore reluctant to
%draw any astrophysical conclusions from this star. Hence, we restrict our
%further analysis to those stars that are expected to oscillate with
%relatively low 
%frequencies, which basically limits the sample to the first five stars in
%Table~\ref{tab1}.
The maximum signal-to-noise ratio (SNR) is 3.8 (in amplitude) between
the highest peak and the average level 
measured in the frequency range 67-87\muhz, denoted $n_{\mathrm{w}}$
in Figure~\ref{f3} for stars 2 -- 7.  Hence, extracting individual peaks
from these spectra is barely justified. Instead we look for broad
envelopes of excess power by smoothing the spectra and compare that to a
Harvey model for the background noise \citep{Harvey85}.
The solid black curve in Figure~\ref{f4} shows the smoothed power spectra
(scaled to power density) and an overlaid one-component Harvey model
(dashed curve) %of the form
\begin{equation}
  p(\nu)=n_\mathrm{w} + \frac{4\sigma^2\tau}{1+(2\pi\nu\tau)^2},
\label{harvey}
\end{equation}
where $p(\nu)$ is the power density at frequency, $\nu$, $n_\mathrm{w}$ is
the lowest level in the power spectrum (assumed to be dominated by white
noise), and $\sigma$ and $\tau$ are the rms amplitude and characteristic
time scale for the background, respectively. Smoothing is done by
convolving the spectra with a Gaussian function of width 4\dnu, which follows
the approach by \citet{Kjeldsen07}.
%maybe abit about that exp=2 fit best and that the data did not support
%fitting more than one Harvey component
For comparison we also show with a dashed curve the average smoothed spectrum of star
No. 9 and 10, which both have expected frequencies well beyond the plotted
range and very low expected amplitudes. Hence, this serves as a reference where we
do not expect any detectable signal from oscillations.
There appears to be excess power in five out of these six stars, 
roughly following the same trend as the estimated \numax\ (horizontal bar). 
%However, the is a tendency that the observed power   
%is located at slightly higher frequencies than predicted, which could 
%suggest either that scaling the solar acoustic cut-off frequency is not a 
%good measure for predicting the frequency of maximum power in metal poor
%giants, or that one or more of the the stellar parameters in
%Table~\ref{tab1} are systematically off. We are more inclined to suspect
%the latter reason.  
%If referee wants details on how we 'fit' the backgound:
%The Havey models were matched by eye, where for the two brightest stars we
%fixed the white noise level to that measured in the spectrum in the range
%67-87\muhz. However, for the fainter stars where there were no flat noise
%plateau this approach did not work and we instead forced/adjusted the
%white noise compoment to make/enforce a resonable match with the lowest
%point in the spectrum just beyond the 87\muhz\ frequncy mark. 
We note that to some extent the location of the excess humps seen in
Figure~\ref{f4} are reproduced by the ``oscillation-quiet'' reference
stars when the same degree of smoothing is applied. In particular, the
spectrum of star No. 8, which was discarded due to its high noise level,
looked very similar to what we see for star No. 6, suggesting that its
excess hump is a noise feature. This is further supported by its location 
at half the orbital frequency (174 $\mu$Hz) of the spacecraft.
%They merely represented noise  
%wiggles/humps characterised by the width of the Gaussian function used to
%smooth the spectrum. Due to the very low signal-to-noise, one could suspect
%that the excess power shown for star 510, 355, and 457 is in fact a result
%of that same effect. 

%Three stars had expected oscillations frequencies near
%200\muhz, which is not vell sampled by our data. While these stars can not
%be persued to look for oscilations we decided to use them as
%reference stars with no expected excess in the investigated frequency range
%0--100\muhz.
%One star (4) showed significantly higher noise in
%the Fourier spectrum than the rest and was omitted. In Figure~\ref{} we
%show the power spectra of the remaining six stars.

\section{Mode lifetimes and amplitudes}\label{amplitudes}
\begin{figure}
%\epsscale{1.0}
\includegraphics{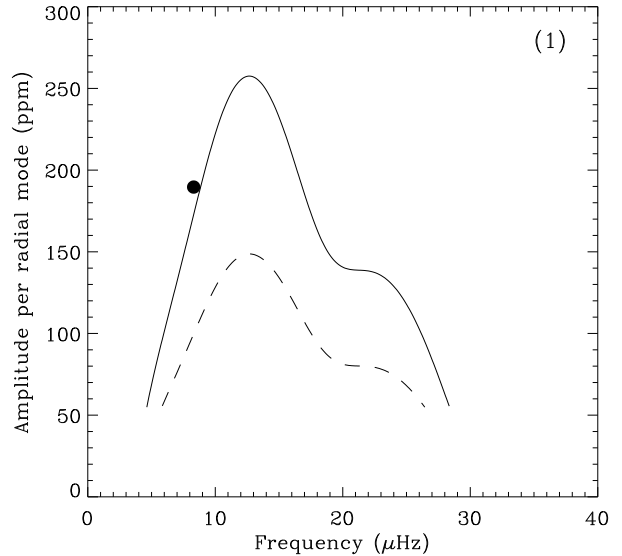}
\caption{Amplitude per radial mode for star No. 1 assuming only radial modes (solid curve)
  and non-radial modes in addition (dashed curve). The dot marks 
  the predicted \numax\ and amplitude from scaling relations
  (Table~\ref{tab1} cols (10) and (13)). 
\label{f5}} 
\end{figure}

In the following we use the results from the previous section to estimate
the amplitude per mode under the assumption the excess power levels seen in 
Figure~\ref{f4} are due to solar-like oscillations.
Due to inadequate time resolution and signal-to-noise of our data, the mode
lifetimes were not attainable in any of the stars.  
Hence, we measure the amplitudes following the approach suggested by
\citet{Kjeldsen05}, which is independent of mode lifetime. First we
subtract the background from the smoothed 
power density spectrum and multiply with the mean mode spacing. As we do
not know whether non-radial modes are excited in these stars, we used the
following average mode spacings to bracket the possible range, \dnu\ if
radial modes dominate the spectrum and \dnu/3 for a Sun-like case with both
radial and non-radial modes taking into account the relative mode
visibilities from whole-disk integrated intensity observations
\citep{Kjeldsen07}. We finally take the square root to
get the amplitude. From Figure~\ref{f5}, which shows the result for star No. 1, 
we conclude that $(L/M)^{0.7}$\teff$^{-2}$ scaling (dot) provides better
agreement with our measured amplitude range than does the
$LM^{-1}$\teff$^{-2}$ scaling. This supports previous findings on
metal-poor stars by \citet{Bedding06} and \citet{Frandsen07}.
We performed similar analyses to the other stars with excess power (No.
2, 3, 6, and 7), which provided ambiguous results. While being in agreement
with $(L/M)^{0.7}$\teff$^{-2}$ scaling, these low levels of excess power were also
consistent with being noise. 

Recently, \citet{Gilliland08} derived an amplitude relation calibrated with
a large set of red giants in the Galactic bulge observed during seven days
with the HST. This relation does not take into account the effect from mode
lifetime or length of the time series. While we in the above analysis
compared observed amplitude per mode with the theoretical predictions, this
relation predicts the rms excess, which \citet{Gilliland08} calculated
relative to `quiet' comparison stars with no detectable oscillation or
granulation signal. Hence,
$\mathrm{rms}_\mathrm{excess}=(\mathrm{rms}^2_\mathrm{star}-\mathrm{rms}^2_\mathrm{quiet})^{0.5}$
is
not directly comparable to the two previously discussed amplitude scaling
relations.  
%While this relation is not generally applicable, 
We have no comparison stars of the same magnitude, and use instead the
first-difference scatter of the star itself to resemble an intrinsic
quiet star, which is similar to calculating the rms of the high-pass
filtered time series. This leads to an rms excess for star No. 1 of
approximately 700 ppm, which is in quite good agreement with the roughly
600 ppm estimated from equation~(5) in \citet{Gilliland08}.

%The Harvey model did not fit the background noise trend, in particular at
%the shoulder at approaximately 20--30\muhz/ for star 344 467 275. 
%We therefore took a pragmatic
%approach and used higher order terms in the exponential in
%equation~\ref{harvey} in order to overcome this problem. Similar simple
%solutions to the same problem have been shown by (Harvey1993,Aigrain2003).

\section{Regular frequency spacings}\label{autocorr}
Further evidence that would support the existence of solar-like
oscillations in these data would be a regular series of
peaks in the power spectrum corresponding to either the large frequency
spacing, \dnu, if the spectrum is dominated by radial modes or half that
if we have significant non-radial modes of degree $l=1$ as well.
Such regular spacing can be revealed by calculating the autocorrelation
function across the power spectrum even for very low signal-to-noise cases
where the excess power is not directly apparent in the power spectra
\citep{Chaplin08}. 
We calculated the autocorrelation functions for all six stars
shown in Figure~\ref{f3}, but found no
compelling evidence of regular spacing that were in agreement with the
expected \dnu. For our most promising star No. 1 the excess power is
confined to such a narrow frequency range that the autocorrelation function
did not provide robust results. Instead we extracted the eight highest
peaks with $\mathrm{SNR}>4$ relative to the white noise level, which hinted
a spacing of about 2.3\muhz~between five consecutive peaks. This
slightly larger than expected frequency spacing is in qualitative agreement
with the observed excess power also being at a slightly higher frequency
range than expected (see Table~\ref{tab1}). However, we are cautious about
claiming any detection of a regular series of frequencies because the
2.3\muhz~spacing is only 2--3 times the general spacing 
seen between successive peaks in the power spectrum, which implies the
spacing is barely resolved, and likely influenced if not possibly caused by
the limiting frequency resolution.

%The star that shows the most significant
%excess of power (No 1) also show a clear peak in the auto correlation
%function at 2.5\muhz. Unfortunately the predicted frequency spacing of
%\dnu$=1.6$\muhz, barely resolvable with our data, and we are reluctant to
%claim any clear detection of a frequency spacing in agreement with
%solar-like oscillations.

%in varous region while giving a weight to each bin 
%the power spectrum. The weight is determined by a Gaussian
%function centered at the expected \numax\ with a width of \numax/2. This
%width follows the relation found empirically from the width of the
%excess power from oscillations in the Sun, the sub giant $\beta\;$Hyi, and
%the late G giant $\xi\;$Hya \citep{Stello07}. We use a high order Gaussian
%function where the exponent equals 8 instead of 2 to obtain a flat
%top and smooth wings, similar to a box convolved by an ordinary Gaussian.
%To get a more `clean' autocorrelation function we set all 
%points in the power spectrum lower than the median power equal to the
%median value before calculating the autocorrelation. 
%MAYBE SHOW ONE EX OF SPECTRUM AND WEIGHT FUNCTION

\section{Discussion}\label{discussion}
The marginal detections of oscillations in this experiment can largely be
attributed to the data acquisition process, which was far from ideal for our
purpose due to the large scale dithering, extremely saturated stars,  
and poor duty cycle.  The lowest noise levels
reached in an amplitude spectrum for these data are
about 50 ppm.  For stars at brightness comparable to our
star No. 1 ($V$ $\sim$ 12) the Kepler Mission \citep{Borucki07}
will provide 30 minute sampling on stars unsaturated in underlying
6 second co-added exposures, with a Poisson noise level of 65 ppm.
With the expectation of near 100\% duty cycle a 23 day interval
will provide amplitude spectrum noise levels of about 4 ppm,
or more than an order of magnitude better than those obtained here.
{\change From the CoRoT mission we have already started to see the first
  glimpse on what to expect from long term monitoring of red giant stars
  using high-precision space photometry \citep{Hekker08,Kallinger08a,Ridder09}.}
Coupled with a near perfect window function and observations 
expected to extend over 3.5 years for some 1000 red giants 
with $V$ $\sim$ 12, the Kepler data will provide several orders
of magnitude gain in capability relative to these serendipitously
available data from a (non-ideal for our purposes) HST program.

\section{Conclusion}\label{conclusions}
We used data obtained with the HST over 23 days to search for solar-like
oscillations in the red giant population of the metal-poor cluster
NGC6397. We found evidence for excess power in the power spectra
in agreement with expected frequencies from solar scaling. 
Except for one star this evidence is
tentative due to the low signal-to-noise, and extraction of individual mode
frequencies was not justified. As a consequence mode lifetimes and mode
identification was not attainable. For the most promising star (No. 1), the most
luminous in our sample, we were able to estimate the amplitude 
per mode from the excess power, which favours that the amplitude follows a
$(L/M)^{0.7}$\teff$^{-2}$ relation rather than a $LM^{-1}$\teff$^{-2}$ relation when scaled from the
Sun. 
%The excess rms (oscillations+granulation) of this star is also in good
%agreement with what have been measured in a large sample of Galactic bulge
%stars. 
Unfortunately, the frequency resolution of our dataset does not allow 
convincing verification of the typical solar-like regular series of
peaks in the power spectrum. However, these results still represent the
best available evidence of solar-like oscillations in such metal-poor
stars.

%(see \S~\ref{stello}--\ref{sec:shotgun}).

\acknowledgments
DS acknowledge support from the Australian Research Council.
RG acknowledges support through GO/AR-11254 from STScI.

%Facilities: \facility{HST(WFC1)}

\clearpage

\bibliography{bib_complete}

\clearpage

\end{document}